# Galaxy Formation and Large-Scale Bias

Guinevere Kauffmann[1], Adi Nusser[2] & Matthias Steinmetz[1]

[1] *Max-Planck Institut für Astrophysik, D-85740 Garching, Germany*
[2] *Institute of Astronomy, Madingley Road, Cambridge, CB3 0HA, United Kingdom*

## Abstract

We outline a simple approach to understanding the physical origin of bias in the distribution of galaxies relative to that of dark matter. The first step is to specify how collapsed, virialized halos of dark matter trace the overall matter distribution. We define the quantity $M_*$ to be the halo mass that has typically just collapsed by the present day. It can then be shown that on large scales, halos of mass $M_*$ are unbiased tracers of the underlying matter distribution. Halos with masses greater than $M_*$ are positively biased, while halos less massive than $M_*$ are antibiased with respect to the dark matter. These conclusions are independent of the assumed shape of the power spectrum. The next step is to make a connection between halos and the luminous galaxies we observe. We appeal to the results of semi-analytic models of galaxy formation that are tuned to fit the observed luminosity functions of local groups and clusters. Using these models, we are able to specify the luminosities and morphological types of the galaxies contained within a halo of given mass at the present day.

We have also used a high-resolution N-body simulation of a cold dark matter (CDM) universe to study the bias relation in more detail. The differences between the galaxy and dark matter distributions are quantified using a number of different clustering statistics, including the power spectrum, the two-point correlation function, the void probability function and the one-point probability density function. We arrive at the following general conclusions:

1. A comparison of the galaxy and dark matter density fields shows that linear biasing is a good description on large scales for galaxies of all types and luminosities.

2. The bias factor $b$ depends on the shape and normalization of the power spectrum. The lower the normalization, the larger the bias. More bias is obtained for spectra with more power on large scales. For "realistic" models, $b$ ranges from 1 to 2.5.

3. Galaxies of different luminosity or morphology have different bias factors.

4. The scale dependence of the bias factor is weak.

# 1 Introduction

One of the most important goals of modern observational cosmology is to measure the mass density parameter $\Omega$. However, many methods for obtaining dynamical estimates of the mass content of the universe depend crucially on how visible galaxies trace the underlying dark matter distribution (see Dekel 1994 for a review). It is likely that galaxies will only form in regions where gas can reach high enough density to cool and form stars. The distribution of galaxies may then be *biased* relative to that of the dark matter. It is often assumed that the galaxy and mass density fields are related through a constant bias factor $\delta_g(x) = b_g \delta_{dm}(x)$. Bardeen et al (1986) demonstrated that such a linear relation arises if galaxies form at the high peaks of the smoothed linear density field. Nevertheless, it may still be argued that the assumption of linear bias is overly simplistic and physically unmotivated.

The validity of any biasing relation can be assessed by comparing the observed galaxy distribution with the mass-density field inferred from the large-scale peculiar motions of galaxies. Under the assumption of linear bias, these comparisons yield the quantity $\beta \equiv \Omega^{0.6}/b_g$. Current analyses give $\beta = 1 \pm 0.4$ (Dekel et al 1993; Hudson et al 1995). On the other hand, a complementary method that compares the observed peculiar motions with the velocities derived from the galaxy density field, favours lower values of $\beta$ (Nusser & Davis 1994, Davis & Nusser 1995). Other indirect estimates of $\beta$ from the distortion of the clustering pattern of galaxies in redshift space yield values in the range 0.4-1 (Hamilton 1995, Cole, Fisher & Weinberg 1995, Fisher, Scharf & Lahav 1994, Heavens & Taylor 1995, Fisher & Nusser 1995). One important question that then arises is whether this scatter in the measured value of $\beta$ can be attributed to deviations from the linear bias relation.

The physical origin of large-scale bias is linked to the complex processes that regulate the formation of a galaxy, such as star formation, supernova feedback and merging (Dekel & Rees 1984). It is a well-known observational fact that different types of objects in the universe can have very different clustering properties. For example, the correlation amplitude of galaxy clusters is a strong function of cluster richness (Bahcall & Soneira 1983; Klypin & Koplyov 1983). Elliptical galaxies are more clustered on small scales than spiral galaxies (Dressler 1980). It has also been speculated that the processes controlling galaxy formation may depend on environment. Efficient galaxy formation in clusters has often been invoked as a means of reconciling the low mass-to-light ratios of these systems with a critical mass density Universe. Clearly it would be valuable to develop a technique for calculating galaxy bias within the framework of a physically realistic theory of galaxy formation.

In the standard scenario of structure formation, a dominant dissipationless component of dark matter is assumed to aggregate by gravitational instability into clumps, the virialized parts of which are usually called dark matter halos. Formation of galaxies by the cooling and condensation of gas within these halos was first considered in a seminal paper by White & Rees (1978). More recently, new techniques have been developed that have permitted more detailed comparisons with observational data to be made (White & Frenk 1991; Lacey et al 1993; Kauffmann, White & Guiderdoni 1993; Cole et al 1994). In parallel with this analytic work, numerical simulations of galaxy formation that include gas as well as dark matter have provided independent confirmation of many of the qualitative aspects of the original White & Rees scenario (Katz & Gunn 1991; Navarro & White 1994; Evrard, Summer & Davis 1994; Steinmetz & Müller 1995). Although many of the detailed physical processes that regulate the formation and evolution of a galaxy, such as star formation and supernova feedback, remain poorly understood, the basic halo-galaxy connection is now theoretically well established.

The first step towards a physically realistic treatment of bias is, therefore, a mathematical description of how the mass function of dark matter halos in a given region of the universe is modulated by the surrounding density field. This modulation is the so-called "natural bias" that has been discussed by many authors (Kaiser (1988), Dekel & Silk (1986), White et al (1987), Cole & Kaiser (1989)). It is simple to show that for a Gaussian initial density field, massive halos will preferentially be located in overdense regions of the universe, whereas low-mass halos will tend to be found in underdense regions. In the first section of this paper, we show how the Press-Schechter theory and its extensions can be used to obtain an estimate of this effect for a specified set of cosmological initial conditions.

The second, perhaps more crucial step, is then to specify how the luminous galaxies that we observe are related to the dark halo distribution. There are three possible approaches to this problem. The most direct and believable way is to determine the halo-galaxy connection observationally by obtaining the luminosity functions of galaxies in virialized systems as a function of their mass. Bright field galaxies like our own Milky Way are thought to reside at the centres of dark matter halos with masses between $10^{12}$ and $10^{13} M_\odot$. Many of these galaxies also have a number of fainter satellite companions. The Milky Way has two prominent irregular-type satellites, the Large and Small Magellanic Clouds. It also has a number of less luminous dwarf elliptical companions. Studies of the satellite galaxies around nearby isolated bright galaxies (Zaritsky et al 1993; Lorrimer et al 1994) suggest that this may be fairly typical. Groups and clusters of galaxies are probably situated in more massive halos. One may use the radial velocities of the galaxies in a group to get a dynamical estimate of its halo mass. These range from $\sim 5 \times 10^{13} M_\odot$ for a small group to $\sim 10^{15} M_\odot$ for a rich cluster like Coma. More accurate mass determinations may be possible in future by making use of techniques based on gravitational lensing (Kaiser & Squires 1993; Seitz & Schneider 1995). There have been some claims that the luminosity function of groups and clusters has a universal form that is well-approximated by a Schechter function. In nearby clusters, the faint-end slope is estimated to lie in the range -1.2 to -2.2 (Binggeli, Sandage & Tamman 1985; Ferguson & Sandage 1988; Huchra 1985), steeper than the slope found for the field galaxy population. The galaxy population in clusters also differs from that in the field, in that the central regions of clusters are composed predominantly of elliptical galaxies, whereas 70-80 percent of galaxies in the field are spirals (Dressler 1980). Clusters also frequently contain a central, highly luminous elliptical galaxy called a cD. Future large galaxy redshift surveys such as the Sloan, which will also include a wealth of photometric, colour and spectroscopic information, will help clarify exactly how the properties of galaxy groups and clusters vary as a function of their mass.

The second approach is to use very large N body + hydrodynamics simulations capable of resolving the formation of galaxies in a region of the universe large enough to be cosmologically interesting (Cen & Ostriker 1992; Katz, Hernquist & Weinberg 1992). More reliable studies will be possible in future using parallel supercomputers.

The third approach, which is the one we adopt here, is to make use of semi-analytic models of galaxy formation that are constrained to provide good fits to the observed luminosity functions of local virialized systems, such as the Milky Way and its satellites, as well as clusters such as Virgo and Coma. These semi-analytic models are able to specify the luminosity functions of galaxies in halos of different mass with minimal computational effort.

Having specified the connection between galaxies and halos, we are in a position to make a quantitative comparison between the distribution of dark matter and that of galaxies. In this paper, we explore two different methods for doing this. One is to use the analytic formulae to derive the *mean* relation between galaxy and dark matter overdensity. This can also be done as a function galaxy luminosity or morphological type. The second method relies on large cosmological N-body simulations that include only a dissipationless dark matter component. Halos identified in these simulations act as spatial markers, telling us where galaxies should be placed. We thus are able to use these simulations, in conjunction with the semi-analytic models, to produce mock galaxy catalogues. We then apply a number of different statistical measures of galaxy clustering, including the correlation function, the void probability function (VPF) and the one-point probability distribution (PDF), to both the galaxies and the dark matter and see how the results differ from one to the other.

## 2 Analytic Models of Halo Bias

### 2.1 Derivation of Halo Bias Using the Press-Schechter Theory

In a recent paper, Mo & White (1995) discuss analytic models of the clustering properties of dark matter halos in considerable detail. Here we adopt their notation and present only a brief derivation of the halo bias formula. We restrict the discussion to an Einstein-de Sitter universe.

In the standard Press-Schechter theory, the matter contained in a spherical region of the universe of comoving radius $R_0$ is assumed to be part of a single collapsed structure by redshift $z$ if its linear overdensity extrapolated forward to that epoch exceeds the critical value $\delta_c = 1.68(1 + z)$. The mass of this collapsed structure will be $M = 4/3\pi\bar{\rho}R_0^3$, where $\bar{\rho}$ is the mean density of the universe. If we assume that the initial overdensity field is Gaussian, the comoving number density of halos, expressed in current units as a function of $M$ and $z$ is

$$n(M,z)dM = -\left(\frac{2}{\pi}\right)^{1/2} \frac{\bar{\rho}}{M} \frac{\delta_c}{\Delta} \frac{d\ln\Delta}{d\ln M} \exp\left[-\frac{\delta_c^2}{2\Delta^2}\right] \frac{dM}{M} \qquad (1)$$

The quantity $\Delta$ is the r.m.s. fluctuation of mass smoothed over a radius R:

$$\Delta^2(R,z) = \int_0^\infty 4\pi k^2 dk |\delta_k(z)|^2 W^2(kR), \qquad (2)$$

where $W(kR)$ is the adopted filter function.

In order to determine how halos are biased relative to the mass, a formula describing the relationship between the mass function of halos and the local density field is required. This is provided by an extension of the Press-Schechter theory due to Bower (1991) and Bond et al. (1991). The fraction of mass in a region of radius $R_0$, mass $M_0$ (corresponding to r.m.s. variance $\Delta_0$) and present extrapolated overdensity $\delta_0$, which at redshift $z1$ (corresponding to extrapolated critical overdensity $\delta_1 = 1.68(1 + z_1)$ ) is in halos of mass in the range $M_1$ to $M_1 + dM_1$ is

$$f(1|0)dM_1 \equiv f(\Delta_1, \delta_1|\Delta_0, \delta_0)dM_1 = \frac{1}{(2\pi)^{1/2}} \frac{\delta_1 - \delta_0}{(\Delta_1^2 - \Delta_0^2)^{3/2}} \exp\left[-\frac{(\delta_1 - \delta_0)^2}{2(\Delta_1^2 - \Delta_0^2)}\right] \frac{d\Delta_1^2}{dM_1} dM_1 \qquad (3)$$

So the average number of $M_1$ halos at redshift $z_1$ in a spherical region with comoving radius $R_0$, overdensity $\delta_0$, is simply

$$N(1|0) = \frac{M_0}{M_1} f(1|0), \qquad (4)$$

By definition, we require that $M_1 \leq M_0$. So by combining equations 1 and 4, we derive the average overdensity of halos with mass $M_1$ identified at redshift $z_1$ in regions $(R_0, \delta_0)$ as

$$\delta_h(1|0) = \frac{N(1|0)}{n(M_1, z_1)V_0} - 1, \qquad (5)$$

where $V_0 = 4/3\pi R_0^3$.

Note that this is the overdensity of halos in Lagrangian space. In order for the formula to be useful in the analysis of real galaxy catalogues, we need to derive the halo bias formula in physical (Eulerian) space. Following Mo & White (1995), we make the simple assumption that a mass shell $(R_0, \delta_0)$ will contract or expand, depending on the value of $\delta_0$, according to the spherical model:

$$\frac{R(z)}{R_0} = \frac{3}{10} \frac{1 - \cos\theta}{\delta_0} \qquad (6)$$

$$(1 + z)^{-1} = \frac{3 \times 6^{2/3}}{20} \frac{(\theta - \sin\theta)^{2/3}}{\delta_0} \qquad (7)$$

One is then able to relate $\delta_0$ to the physical mass overdensity $\delta \equiv [(R_0/R(z))^3 - 1]$. measured in a sphere of radius $R$ at redshift $z$. Mo & White also show that in the limit where $R_0 \gg R_1$, and $|\delta_0| < \delta_1$, equation 5 reduces to the peak-background split approximation discussed by Bardeen et al (1986) and Cole & Kaiser (1989). In this approximation, the effect of long-wave perturbations in the density field is modelled as a constant shift in the background density. This leads to a modulation in the collapse times of non-linear objects; the non-linear condensations in a perturbed region are identical to those seen in an unperturbed region, but at a slightly different time.

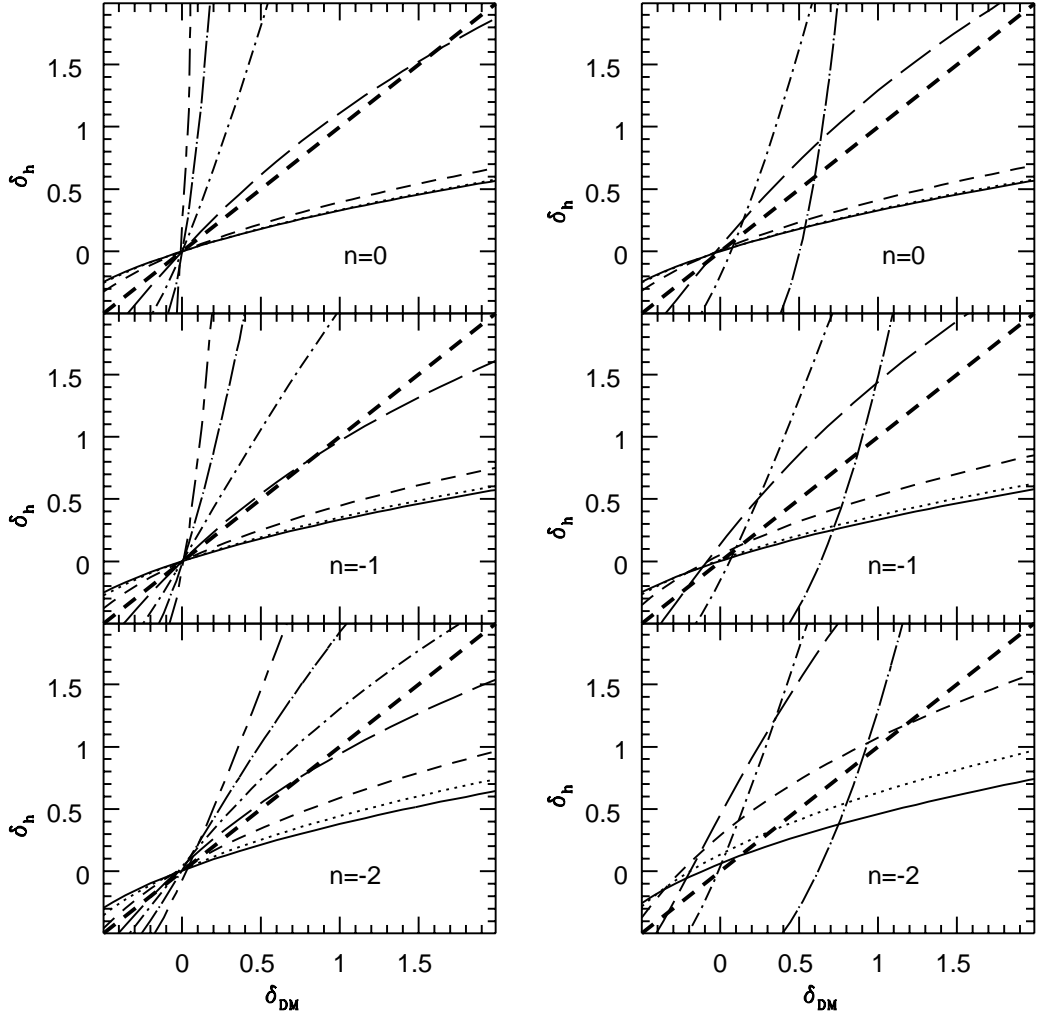

Figure 1: The overdensity of halos in spheres containing mass 1000 $M_*$ is plotted against dark matter overdensity in the left column. The right column is for spheres containing mass 20 $M_*$. Results are shown for power-law spectra with indices $n = 0, -1$ and $-2$. Solid, dotted, short-dashed, long-dashed, short dashed-dotted, long dashed-dotted and short dashed- long dashed lines show the bias relations for halos with masses 0.001, 0.01, 0.1, 1, 3, 10 and 30 $M_*$, respectively. The thick dashed line is the "unbiased" relation $\delta(\text{halos}) = \delta\,(\text{DM})$.

It is now useful to define a quantity $M_*$ at a redshift z, so that

$$\Delta(M_*) = \delta_1 = \delta_c(1+z), \tag{8}$$

i.e. $M_*$ is the mass of the halo that is typically just collapsing at redshift $z$. In figure 1, we show halo bias relations for power-law initial density fluctuations. Halos are identified at $z = 0$. We plot the halo overdensity versus the dark matter overdensity evaluated in spheres containing mass $M_0 = 1000 M_*$ in the left hand column, and in spheres containing mass $20 M_*$ is the right hand column. The thick dashed line in the plot shows the "unbiased" relation, $\delta_h = \delta_{dm}$. The series of thin lines in each plot shows the bias relation for halos of different mass. As can be seen, halos with masses less than $M_*$ are *antibiased* with respect to the dark matter, i.e. they occur preferentially in underdense regions of the universe. Halos with masses greater than $M_*$ are positively biased with respect to the dark matter. The degree of this bias is dependent on the power spectrum: in spheres of $1000 M_*$, massive halos exhibit the most bias for the $n = 0$ case and the least bias for the $n = -2$ case. This may be understood by considering the Press-Schechter formula, which tells us that the fraction of the total mass in the universe contained in halos of mass $M$ is much more steeply peaked around $M_*$ for a power spectrum with $n = 0$ than for a spectrum with $n = -2$. An "average" volume of the universe thus contains halos that span only a narrow range in mass for $n = 0$, so halos much more massive than $M_*$ exhibit more bias. It is interesting to note that in the large spheres, halos with masses near $M_*$ follow the unbiased relation over a substantial range in overdensity, independent of the value of $n$. In addition, the bias relation for halos of all masses is approximately *linear*. This is the limit where the peak-background split is a good approximation. In the small spheres, where $M_0$ is not very much larger than $M_*$, the bias relation is no longer linear for halos with masses comparable to the mass contained in the sphere.

Before one can go ahead and apply these analytic results, it is important to confirm that they do in fact agree with the halo bias relations derived from N-body simulations. Extensive comparisons with simulation results are described in Mo& White (1995). These authors find good agreement between the analytic predictions and the simulation results over a wide range of sphere sizes and over almost the entire range of overdensities probed by the simulations.

## 2.2 Specifying the Halo-Galaxy Connection

We use semi-analytic models of galaxy formation to derive luminosity functions for halos of different mass. These models are described in detail in Kauffmann, White & Guiderdoni (1993, hereafter KWG). Here we present only a brief summary of the main points.

1. **The Merging History of Dark Matter Halos**. Using a technique based on the Press-Schechter theory and its extensions, Kauffmann & White (1993) demonstrated that it was possible to construct Monte Carlo merging history "trees" for present-day halos of given mass. These merging trees allow one to trace the history of each halo mass element through all progenitors from which it formed.

2. **Cooling of Gas in the Halos**. The treatment of gas cooling is based on a model by White & Frenk (1991). A dark matter halo is modelled as an isothermal sphere, truncated at its virial radius, defined as the radius within which the mean overdensity is 200. It is assumed that as a the halo forms, the gas initially relaxes to a distribution that exactly parallels that of the dark matter. The gas temperature is then given by the equation of hydrostatic equilibrium. The cooling radius is defined as the radius within the halo at which the cooling time is equal to the Hubble time. If the cooling radius lies outside the virial radius, all infalling gas cools immediately. If the cooling radius lies inside the virial radius, the cooling rate is determined by a simple inflow equation.

3. **Star Formation and Supernova Feedback**. Once gas cools, it will fall towards the centre of the halo, forming a dense core where stars will be able to form. The star formation law we choose is given by the simple equation $\dot{M}_{stars} = \alpha M_{cold}/t_{dyn}$, where $M_{cold}$ is the total mass of cold gas in the galaxy, $t_{dyn}$ is the dynamical timescale of the galaxy and $\alpha$ is an adjustable parameter. Stars form with the standard initial mass function given by Scalo (1986). The number of supernovae that are then expected per solar mass of stars formed is $4 \times 10^{-3}$. The kinetic energy of the ejecta of each supernova is $10^{51}$ ergs. We assume that a fraction $\epsilon$ of this energy goes to reheat cold gas to the virial temperature of the halo, where $\epsilon$ is taken to be a free parameter.

4. **Merging of Galaxies**. N-body simulations show that when two dark matter halos merge, all substructure is erased on a very short timescale. However, the dense baryonic cores of the halos may survive for a considerably longer period of time. Navarro, Frenk & White (1995) have shown that in simulations of the formation of a $\sim 10^{12} M_\odot$ halo, the baryonic cores coalesce on a timescale that is well-approximated by the time required for dynamical friction to erode the orbit of the smaller "satellite" system, causing it to spiral in to the centre of the larger halo and merge. In the semi-analytic models, the timescale over which a satellite galaxy will merge is set by the dynamical friction time
$$t_{dynf} = \frac{1.17 V_c r_c^2}{\ln \Lambda G M_{sat}}, \qquad (9)$$
where $V_c$ is the circular velocity of the primary halo, $r_c$ is the virial radius of the halo, $M_{sat}$ is the baryonic + dark matter mass of the orbiting satellite and $\ln \Lambda$ is the usual Coulomb logarithm.

5. **Determining Galaxy Morphology**. It is assumed that the cooling and subsequent infall of gas results in the formation of a rotationally supported disk galaxy at the centre of the halo. The merging of two roughly equal mass disks forms a spheroidal merger remnant. If more gas can cool onto this merger remnant, a spiral galaxy consisting of both a disk and a bulge component is produced. If no further cooling occurs, for example if the galaxy is accreted by a larger halo, then the merger results in the formation of an elliptical galaxy. Galaxies are classified into early-type and late-type systems according to their stellar bulge-to-disk ratios as given in KWG.

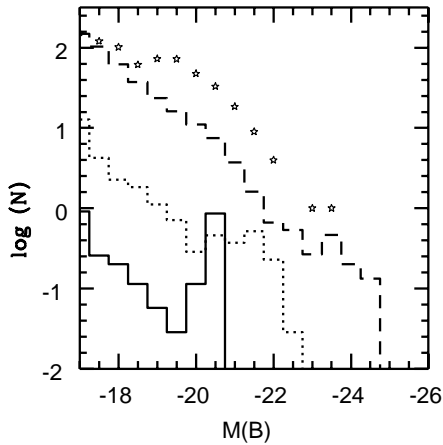

Figure 2: The B-band luminosity function of galaxies contained in halos of 3 different masses. The solid line is a "Milky-Way" halo of mass $5 \times 10^{12}$ M$_\odot$. The dotted line is a galaxy "group" halo of mass $5 \times 10^{13}$ M$_\odot$. The dashed line is a "cluster" halo of mass $10^{15}$ M$_\odot$. The stars show the luminosity function of the Coma cluster from Thompson & Gregory (1992).

6. **Tranformation to Observables**. The spectrophotometric models of Bruzual & Charlot (1993) are used to transform the star formation history of each galaxy into present-day estimates of magnitude, colour, etc.

7. **Fixing the Free Parameters**. We adopt a baryon density, $\Omega_b$, of 0.1 for a Hubble constant $H_0 = 50$ km s$^{-1}$ Mpc$^{-1}$. The free parameters $\alpha$ and $\epsilon$ that control the star formation rate and supernova feedback efficiency are fixed by requiring that the central galaxy of a dark matter halo of mass $5 \times 10^{12} M_\odot$, circular velocity 220 km s$^{-1}$, have luminosity and gas content similar to that of our own Milky Way.

The galaxy formation model described above is the one we consider to be most realistic. It will be referred to as model R in future. We also explore the effect of changing some of the parameters of the model. In model NM, satellite galaxies never merge. This model maximises the number of individual galaxies within one dark matter halo. In model HM, the merging timescale is artificially reduced by a factor 20. Finally, in model F, we increase the value of $\Omega_b$ to 0.2. More gas is then able to cool and the efficiency of supernova feedback must increase in order to compensate and produce a "Milky Way" of the same luminosity as before. Note that all the models are normalized to the Milky-Way as described above.

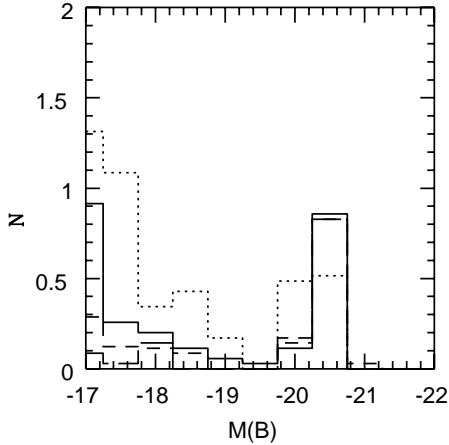

Figure 3: The B-band luminosity function of galaxies contained in a Milky-Way halo of mass $5 \times 10^{12}$ M$_\odot$ for the 4 different variations in galaxy formation parameters described in the text. The solid line is model RB (the realistic model). The dotted line is model NM (no merging). The short dashed line is model F (high feedback). The long dashed line is model HM (high merging).

The B-band luminosity function of galaxies contained in halos of three different masses is shown in figure 2 for model R. The solid line is the luminosity function of a $5 \times 10^{12} M_\odot$ halo, which is the mass of a Milky-Way system. These halos contain one central galaxy with $M(B) \sim -20$ and an average of 1.6 satellites brighter than $M(B) = -17$. This agrees quite well with the observations. The dotted line in the plot is the luminosity function of a halo with mass $5 \times 10^{13} M_\odot$, the mass of a typical galaxy group. These halos typically contain 25 galaxies down to a limiting magnitude of -17. Finally, the dashed line is the luminosity function of a rich cluster halo of mass $10^{15} M_\odot$, which contains 400 galaxies brighter than $M(B) = -17$. Note that the shapes of the three luminosity functions in figure 2 are very similar, but they are are closer to power-law than Schechter, as there is no clear "break" at any luminosity. The luminosity of the brightest galaxy contained in the halo also increases as a function of the halo mass. The most luminous galaxies in the universe are formed in cooling flows at the centres of clusters, where the accretion of infalling gas and satellites can push the luminosity as high as 100 $L_*$. The total *number* of galaxies contained in a halo scales roughly in proportion to the mass of the halo.

If we compare our cluster luminosity function in figure 2 with the luminosity function of Coma (Thompson & Gregory 1993), we find a good match at the bright and faint ends, but a deficit of galaxies with magnitudes between -20 and -22. The Coma cluster luminosity function has a pronounced "bend" at these magnitudes, whereas our luminosity function is well-represented by a single power law. We find a similar discrepancy if we compare the luminosity functions of somewhat less massive halos ( $V_c$= 1000 km s$^{-1}$ corresponding to $5 \times 10^{14} M_\odot$) with the luminosity function of the Virgo Cluster (Binggeli, Sandage & Tamman 1985). This deficiency of bright galaxies translates into cluster mass-to-light ratios a factor of 2 higher than the observed values. We have experimented with the parameters in our models and find that it is not possible to change the number of galaxies in this magnitude range by more than 20-30 percent without simultaneously destroying the good fit at the faint end of the luminosity function. Apparently, power-luminosity functions are a generic prediction of the models. One possibility is that the brightest, most massive galaxies may occur preferentially at the centres of clusters. The observed bend at $L_*$ might simply then arise if the cluster luminosity function has not been evaluated all the way out to the virial radius, which is 3 Mpc for Coma. For the purposes of this paper, we sidestep this problem and artificially *boost* the number of bright galaxies in clusters so that we match the observations. We increase the number of galaxies in the magnitude range -20 to -22 by a factor two in all halos with masses greater than $10^{14} M_\odot$. This model is now referred to as model RB (realistic-boosted).

For comparison, luminosity function of galaxies contained in halos of mass $5 \times 10^{12} M_\odot$ are shown in figure 3 for models RB, NM, HM and F. The main effect is to produce a change in the *slope* of the luminosity function. Merging reduces the number of satellite galaxies and increases the luminosity of the central galaxy. Efficient feedback suppresses star formation in dwarf galaxies. When the models are then renormalized to match the Milky-Way luminosity, the effect is simply to change the luminosity function of the satellites.

## 2.3 Analytic Estimates of Galaxy Bias

We are now in a position to combine the analysis from the previous two sections and derive an analytic estimate of galaxy bias. Results are shown in figure 4 for galaxy formation model RB. We plot the average overdensity of galaxies measured in spheres of physical radius 10 $h^{-1}$ Mpc versus the average overdensity of dark matter in these spheres. We show results for a CDM power spectrum with $\Omega = 1$, $H_0 = 50$ km s$^{-1}$, and two choices of the normalization parameter $\sigma_8$. We also explore the effect of increasing the amount of large-scale power relative to "standard" CDM. The power spectra of $\Omega = 1$, h=0.5 CDM-dominated universes in which one species of massive neutrino contributes $\Omega_\nu$ in the range 0.2-0.3, can be fitted with a "shape"-parameter $\Gamma \sim 0.2$ (Efstathiou, Bond & White 1992). The third panel in figure 4 shows a $\Gamma = 0.2$ model that is normalized to $\sigma_8 = 0.4$, consistent with the COBE measurement of the microwave background fluctutations on large scales. Another possibility that is compatible with both the COBE measurement and the clustering amplitude of galaxies on large-scales is a low-density ($\Omega = 0.2$) CDM model with $H_0 = 100$ km s$^{-1}$ and $\sigma_8 \sim 1$. The analytic formulae derived in section 2.1 can be extended to the case $\Omega < 1$ very easily and results for this model are shown in the fourth panel of figure 4.

The thick line in each plot is the relation expected if galaxies are unbiased tracers of the dark matter. The thin lines show bias relations for galaxies selected at different limiting absolute magnitudes. The main effect of changing the parameter $\sigma_8$ is to change the value of $M_*(0)$, the mass of the halo that has typically just collapsed by redshift 0. For $\sigma_8 = 1$, $M_*\sim 10^{14}M_\odot$, the mass of a galaxy cluster. For $\sigma_8=0.4$, $M_*\sim 10^{13}M_\odot$, the mass of a Milky Way galaxy halo. As we discussed in section 2.1, galaxies that occur predominantly is halos more massive than $M_*$ will be biased tracers of the mass, whereas those that occur in halos less massive than $M_*$ will appear antibiased. This explains why in figure 4, the brightest galaxies, which occur predominantly in massive groups and clusters, are biased relative to the mass in models with low normalization, but not in the more highly clustered models. The models with more large-scale power show significantly more bias than the "standard" CDM models with the same normalization, because the change in shape parameter increases the amount of power on large scales, but reduces the amount of power on galaxy and cluster-sized scales, and hence lowers the value of $M_*$. Another conclusion one can glean from figure 4, is that large-scale luminosity segregation is only expected in the models that have low normalization. For $\sigma_8=1$, galaxies of all luminosities trace the matter distribution in much the same way.

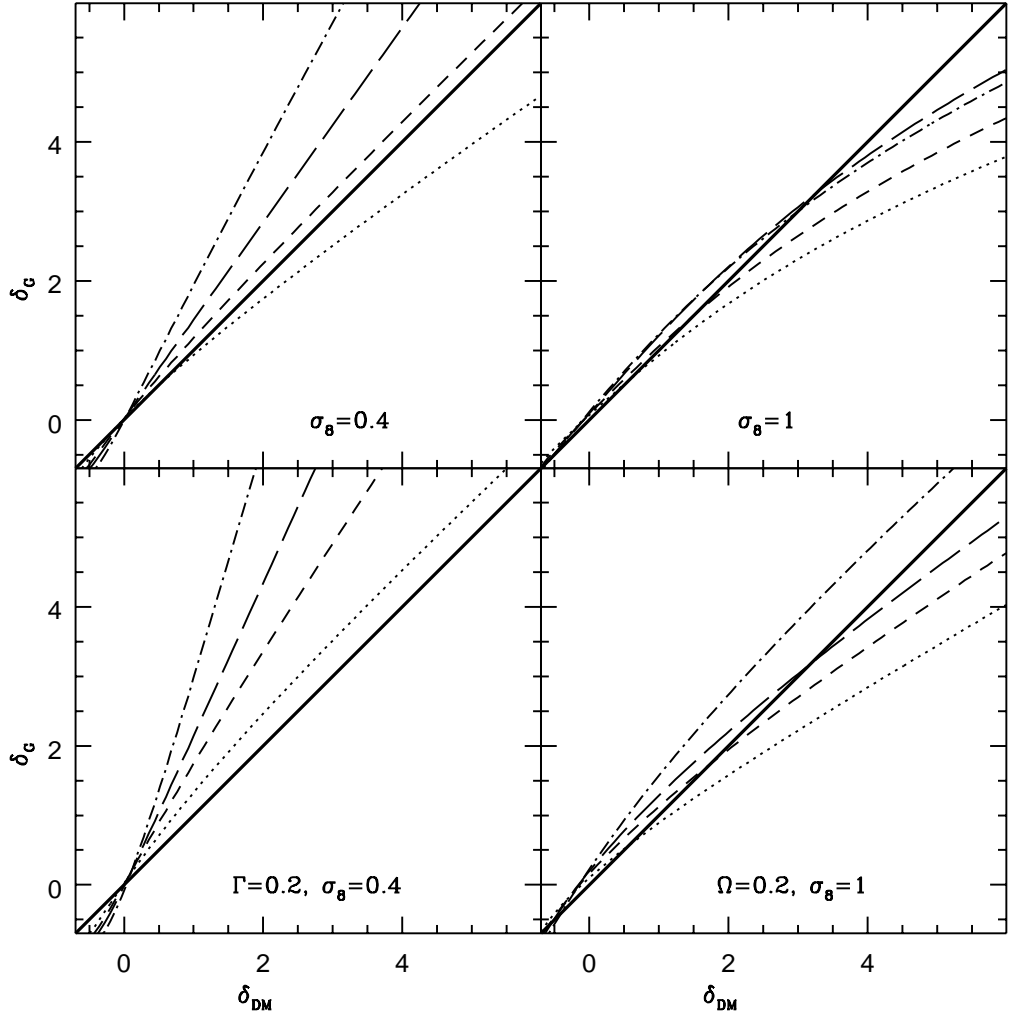

Figure 4: The overdensity of galaxies in spheres of radius 10 h$^{-1}$ Mpc is plotted against dark matter overdensity. Results are shown for 2 different normalizations of the CDM power spectrum, as well as a Cold+Hot dark matter-type spectrum ($\Omega = 1, h = 0.5, \Gamma = 0.2, \sigma_8 = 0.4$) and a low-density CDM spectrum ($\Omega = 0.2, h = 1, \sigma_8 = 1$). Dotted, short dashed, long dashed, and short dashed-dotted lines are for galaxies selected to be brighter than M(B)= =-18, -19, -20, and -21, respectively. The thick solid line shows the unbiased relation $\delta$(galaxies)=$\delta$(DM).

In figure 5, we show galaxy bias relations for the same set of models, but this time evaluated in spheres of radius 4 h$^{-1}$ Mpc. It should be noted that both in figure 4 and figure 5, a *linear bias relation* is a good approximation over a significant range in overdensity. We make a linear fit $\delta_g = b\delta_{dm}$ over a restricted range in $\delta_{dm}$ from -0.5 to 1.5 and plot the derived linear bias factor $b$ as a function of smoothing radius $R$ in figure 6. As can be seen, the bias factor $b$ varies only weakly with scale and is approximately constant for smoothing radii greater than 8 Mpc. The derived values of $b$ range from just under 1 for faint galaxies in the $\sigma_8=1$ CDM models, to 2.5 for $L_*$ galaxies in the $\sigma_8=0.4$, $\Gamma=0.2$ model. For optical galaxies with luminosities $\sim L_*$, the rms variance of the galaxy density in 8 h$^{-1}$ Mpc spheres is $\sim 1$, so for consistency we require that $b(L_*)\sigma_8 = 1$. The models that satisfy this requirement are the COBE-normalized CDM model, the $\Gamma=0.2$ model, and the low-density CDM model. CDM models with $\Omega=1$ and low values of $\sigma_8$ do not result in enough bias to explain the observed clustering amplitude of bright galaxies.

Bias is dependent on morphological type as well as galaxy luminosity. This is illustrated in figure 7, where we plot the bias relation separately for early-type and late-type galaxies. Results are shown for the $\sigma_8=0.4$ CDM model. The distribution of galaxies of different type is very similar in underdense regions, but differs markedly in overdense regions where early-type galaxies are more strongly biased than late-types. We would expect to find a similar difference if we split the galaxy population by colour or present-day star formation rate. This is because the cluster population is dominated by red, early-type galaxies that have typically undergone no recent star formation. The field, on the other hand, is dominated by blue, star-forming spiral and irregular galaxies (see KWG for more details).

The main conclusion, therefore, is that the galaxy-dark matter bias relation depends on both the shape and normalization of the initial power spectrum of density fluctuations and the properties of the sample of galaxies one uses to trace the dark matter. It is thus important to ensure that the tracer galaxies constitute a uniform sample: for example, in flux-limited catalogues, more distant galaxies will be brighter on average than galaxies closer to us. If no correction is made for this effect, one may not be sampling the underlying matter distribution in the same way at all points in the survey.

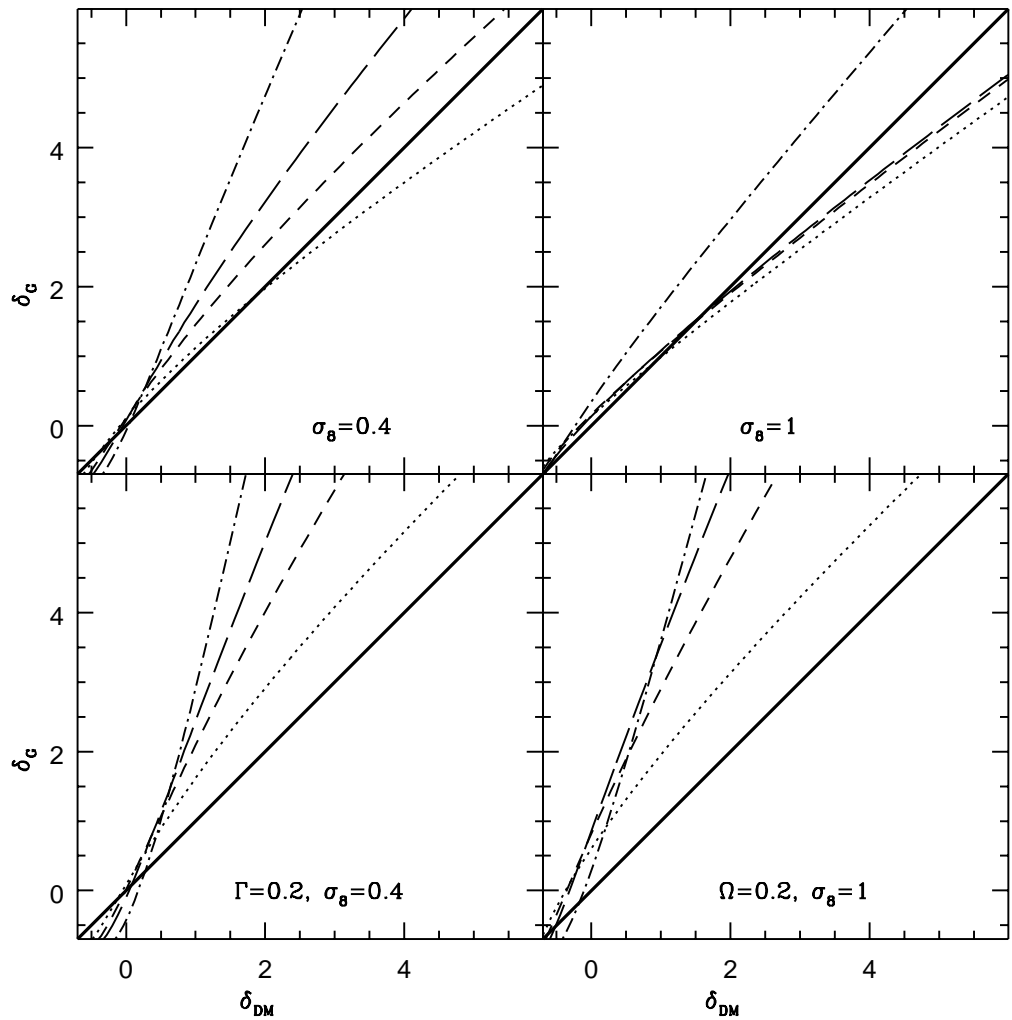

Figure 5: As in figure 4, except that overdensities are evaluated within spheres of radius 4 $h^{-1}$ Mpc.

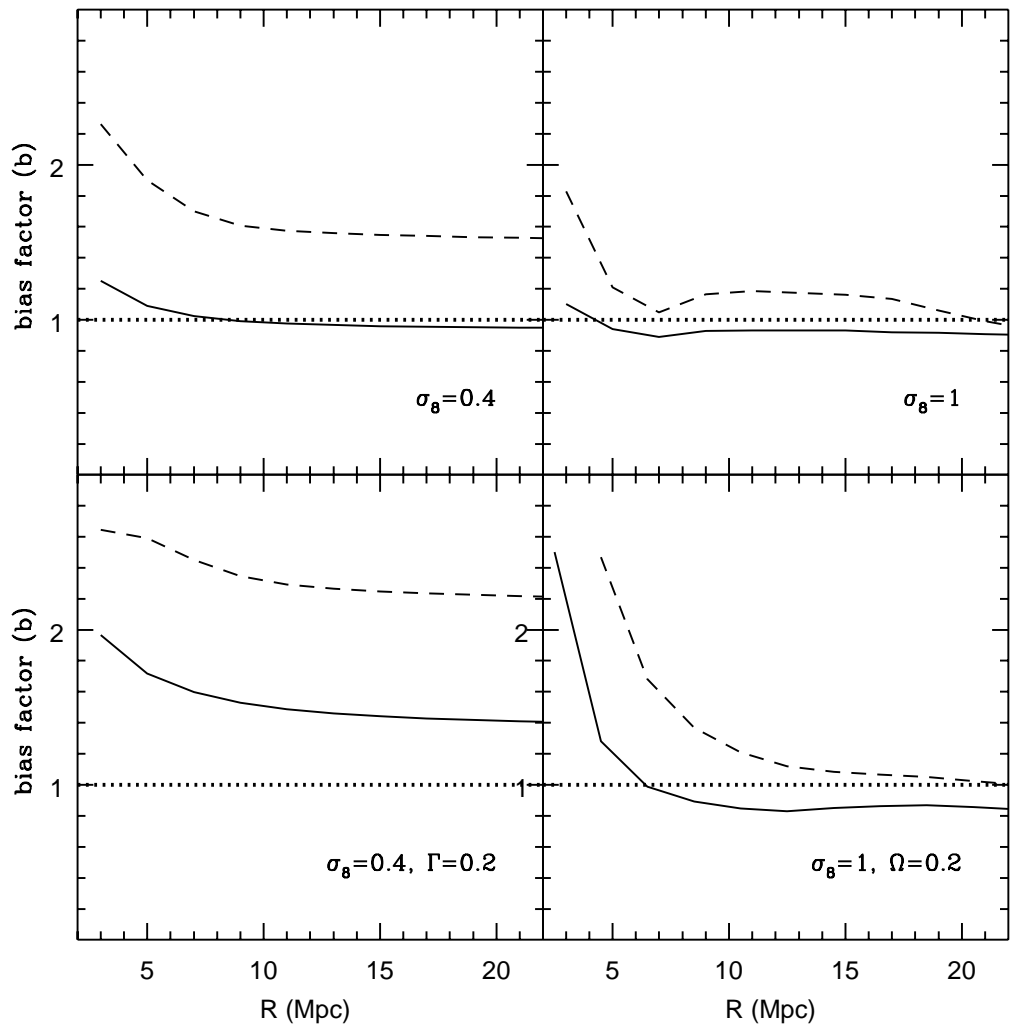

Figure 6: The "linear" bias factor as a function of smoothing radius. The solid line is for galaxies that have absolute magnitude $M(B) \leq -18$. The dashed line is for galaxies with $M(B) \leq -20$.

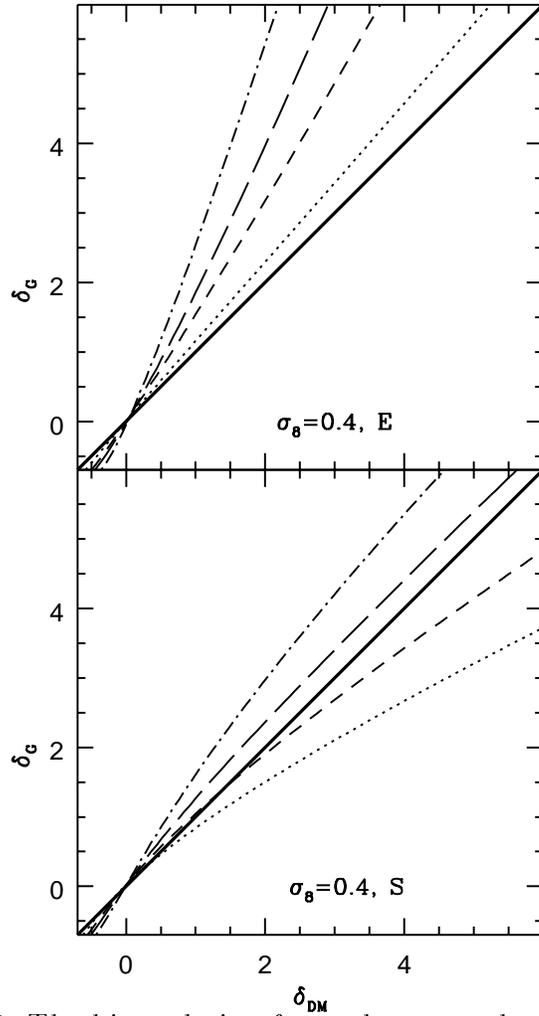

Figure 7: The bias relation for early-type galaxies (top panel) is compared with the bias relation for late-type galaxies (bottom panel) for a CDM spectrum with $\sigma_8 = 0.4$. Explanation of the line types is as given in figure 4.

# 3   Galaxy Bias from N-body Simulations

The analytic models described in the previous section allow one to predict how galaxies of a given luminosity are biased relative to the dark matter within spheres of fixed radius. Large-scale structure theorists probe the clustering properties of galaxies using a wide variety of statistical methods, including two-point and higher order correlation functions, galaxy counts in cells, void probabilities and topology, to name just a few. Many of these clustering statistics are not amenable to a simple analytic treatment.

One way around these difficulties is to use large N-body simulations to construct catalogues of galaxies that can be analyzed in exactly the same way as the real data. This technique has been used many times in the past, but there has been no well-motivated prescription for how galaxies should be identified in a simulation consisting of only dissipationless dark matter (Summers, Davis & Evrard 1995). When two dark matter halos merge, all substructure is erased over a very short timescale. On the other hand, the baryonic cores of the halos can probably survive merging for much longer. It is thus incorrect to make a one-to-one mapping between galaxies and dark halos. Another possibility is to assume that galaxies correspond to peaks in the *initial* linear density field. However, the correspondence between peaks in the initial conditions and collapsed structures at the present day is not always reliable (Katz, Quinn & Gelb 1993 ). In this paper, we make use of our semi-analytic models to assign galaxies to halos identified in the simulations. Our models specify the full luminosity function of galaxies contained within a halo of given mass. We thus bypass the "overmerging" problem inherent in the simulations.

## 3.1   Description of the Simulations

We use the Couchman $P^3M$ code (Couchman 1991) with $128^3$ particles on a $256^3$ mesh to run an $\Omega = 1$ cold dark matter simulation in a box of comoving length 128 Mpc ($H_0 = 50$ km s$^{-1}$ Mpc$^{-1}$). The softening length adopted is 100 kpc. The simulation is normalized to COBE at the final output time, but we also keep less evolved outputs corresponding to $\sigma_8 = 0.67, 0.5$ and $0.4$. Each particle in the simulation has mass $6.89 \times 10^{10} M_\odot$.

The main information we wish to extract from the simulation is the mass and spatial coordinates of all the dark matter halos contained in the box. We identify halos using a friends-of-friends algorithm with a link length of 0.2 times the mean interparticle separation. We then proceed through the list of halos, finding the most bound particle in each one, labelling this particle as the halo centre and evaluating the overdensity within spheres of increasing radius from the centre. The mass assigned to the halo is the mass contained within the sphere of overdensity 200, as stipulated by the Press-Schechter formalism.

One important concern is that the simulation must be able to follow the formation of halos containing only a small number of particles. We do not need to be able to specify the detailed structures of halos, but we do need to be able to estimate their positions and masses accurately. We observe that if too large a timestep is adopted, many of the smaller halos fail to form altogether. We have carried out a series of test runs on a $32^3$ particle simulation using an increasing number timesteps and we have looked for convergence in the number of small halos identified by the group-finding algorithm. Using the results of these tests, we estimate that 500 timesteps (for a time coordinate $p = a^{1.5}$) are required to model the formation of halos consisting of 10 particles. Thus the smallest halos we consider in the analysis have masses $\sim 5 \times 10^{11} M_\odot$. This means we will only be able to resolve galaxies with luminosities greater than $M(B) = -18$, the luminosity of the Large Magellanic Cloud.

Figure 8: A 15 Mpc slice through the simulation volume. The top left panel shows all the dark matter particles. The top right panel shows all the galaxies brighter than $M(B) = -18$. The lower left panel shows all the early-type galaxies in a subvolume of the simulation (outlined by a box in the panels above). The lower right panel shows all the late-type galaxies in this subvolume.

## 3.2 Constructing the Galaxy Catalogue

As described in section 2.2, semi-analytic models of galaxy formation may be used to derive the distribution of the luminosities, morphologies and colours of galaxies contained in a halo of given mass. We run a grid of these models, with halo masses ranging from $10^{11}$ to $5 \times 10^{15} M_\odot$. We store 50 Monte Carlo realizations of the galaxies formed in each halo and select randomly from these realizations. In this way, we make sure we obtain a true measure of the scatter inherent in the galaxy-dark matter bias relation.

Each halo in the simulation is assigned a list of galaxies according to its mass. Each galaxy is then given a position by selecting a halo particle at random and assigning the galaxy its coordinates from the simulation. The galaxy distribution thus traces the dark matter distribution within the virialized regions of halos. This is probably not true in practice, since processes like dynamical friction may act to concentrate galaxies at the centres of halos. We ignore these complications and restrict our analysis to scales larger than a few Mpc, where this is not an important consideration.

Figure 8 shows a slice through the simulations of thickness 15 Mpc at an output time corresponding to $\sigma_8 = 0.4$. Dark matter particles are plotted in the upper left panel and all the galaxies down to an absolute magnitude limit of $-18$ are shown in the upper right panel. The bottom two panels in the figure show the difference in the distribution of early-type and late-type galaxies in a subsection of the slice. Early-type galaxies occur primarily in concentrated knots, whereas late-types are spread out more evenly.

## 3.3 The Galaxy-Dark Matter Bias Relation from the Simulations

In what follows, we show results only for the $\sigma_8 = 1$ and $\sigma_8 = 0.4$ CDM models (note that this the *linear* $\sigma_8$). The bias relation is very different for these two cases. As shown in figure 5, for the $\sigma_8 = 1$ model, galaxies of all luminosities are nearly unbiased tracers of the mass at low dark matter overdensities. For the $\sigma_8 = 0.4$ model, galaxies fainter than

Figure 9: Scatterplots of the overdensity of galaxies versus the overdensity of dark matter evaluated directly from the simulations. Results are shown for CDM power spectra with normalization $\sigma_8 = 1$ and $\sigma_8 = 0.4$.

Figure 10: Scatter plots of the overdensity of galaxies versus the overdensity of dark matter for the $\sigma_8 = 0.4$ CDM model. The four different panels show the results for the four different variations in galaxy formation parameters: RB( realistic), F(high feedback), HM (high merging) and NM (no merging).

$M(B) \sim -19$ are unbiased. Galaxies brighter than -19 exhibit positive bias and the degree of this bias is strongly luminosity-dependent.

We now evaluate the overdensity of galaxies versus the overdensity of dark matter directly from the simulations. We first assign both the galaxies and the dark matter to a $16^3$ grid using the Cloud-in-Cell (CIC) algorithm. These grids are then convolved with a Gaussian filter with smoothing length 1000 km s$^{-1}$ (20 Mpc). Scatterplots of $\delta$(galaxies) versus $\delta$(dark matter) are shown in figure 9 for the two models. As can be seen, our results agree well with the analytic predictions. The relation between $\delta$(galaxies) and $\delta$(dark matter) is seen to be linear. The scatter in the bias relation increases markedly for the bright galaxies as there are fewer of these objects in the simulation volume.

In figure 10, we again show the bias relation for the $\sigma_8 = 0.4$ CDM model, but this time we make a comparison between galaxy formation models NM,HM, F, and RB. As is seen, the choice of galaxy formation model makes only a small difference to the predicted amplitude of the bias. Changing the parameters that control star formation, feedback and merging rates does *not* alter our conclusion that our physically-motivated galaxy formation scheme produces a linear bias relation.

## 3.4 Clustering Statistics

We now apply a number of different statistical measures of clustering to both the galaxies and the dark matter. The purpose of this exercise is to see whether these statistics are sensitive to the differences in the way the galaxies are clustered relative to the underlying mass. If they are, they cannot be used as probes of cosmological initial conditions unless one can come up with a robust way of correcting for these differences. We also construct dark matter catalogues that are randomly sampled from the full particle distribution at the same density as each of our galaxy catalogues. The variations between these random samples provide an estimate of the effect of shot noise on our results.

1. **Correlation Functions**. The two-point correlation function $\xi(R)$ and its Fourier transform, the power spectrum $<|\delta_k|^2>$, are the most widely used statistical measures of galaxy clustering because they are both theoretically fundamental and observationally accessible. One major goal in large-scale structure is to use measurements of the correlations between galaxies on small scales, together with the measured anisotropy of the microwave background radiation on large scales, to determine both the shape and the amplitude of the initial spectrum of density fluctuations. It is thus important to understand how the correlation function measured for galaxies may differ from that measured for the mass.

Correlation functions for the $\sigma_8 = 1$ and $\sigma_8 = 0.4$ models are shown in figure 11. We plot $\xi(R)$ over a range of scales from 2 to 25 Mpc for galaxies selected at different limiting absolute magnitudes. The correlation functions of the corresponding dark matter catalogues are shown as thin lines on the diagram. In the $\sigma_8 = 0.4$ model, the amplitude of $\xi(R)$ for galaxies brighter than $-20$ is boosted relative to that for mass, wheras $\xi(R)$ for fainter galaxies is the same as that for the mass. It is interesting that evidence for a luminosity-dependent clustering amplitude has recently been claimed by a number of authors, including Loveday et al (1995) for the APM-Stromlo survey, Park at al (1994) for the CfA survey and Maurogordato for the SSRS survey (private communication). There is no luminosity-dependent clustering evident for the $\sigma_8 = 1$ model. Figure 12 shows power spectra for the random catalogues and the galaxies. Under the assumption of linear bias, the bias factor $b$ may be computed directly from the galaxy and dark matter power spectra and is shown in figure 13. We assume that the shot noise contribution to the galaxy power spectrum is the same as the contribution to the random dark matter catalogues. Thus by subtracting the two, we obtain an estimate of $b$. As can be seen, the bias factors in figure 13 agree rather well with the analytic predictions of figure 6.

2. **Void Probability Functions**. Voids are an obvious and important feature of the observed galaxy distribution. A simple measure of the size and frequency of voids is the void probability function (VPF), the probability $P_0(R)$ that a randomly placed sphere of radius R contains no galaxies. As shown by White (1979), the VPF is sensitive to the entire hierarchy of n-point correlation functions. In figure 14, we show VPFs for our galaxy and dark matter catalogues. Note that as the limiting magnitude becomes fainter and the galaxy samples get larger, the difference between the galaxy VPF and the dark matter VPF *increases*. The statistic finds no difference between galaxy and dark matter voids for sparse samples of bright galaxies. This tells us that as we proceed down the luminosity function, faint galaxies fill in the voids more slowly than randomly selected dark matter particles.

Weinberg & Cole (1992) demonstrate that in the absence of biasing, the VPF discriminates between Gaussian and non-Gaussian initial conditions for any power spectrum. Non-Gaussian initial conditions tend to affect the void probability function at large radius. Our results show that it will difficult to interpret VPF analyses in practice, because the characteristics of the galaxy survey influence the frequency of the largest voids as well.

3. **The One-Point Probability Density Distribution** As well as considering galaxies as point distributions, one can also view galaxies as Poisson tracers of a continuous underlying field. Convolving the spatial distribution of galaxies with a window function yields a smoothed galaxy density field, and one can apply statistics that measure the properties of this smoothed field. The most basic of these properties is the one-point probability distribution function $P(\delta)$ (or PDF). In the linear regime, $P(\delta)$ provides a meausure of the Gaussianity of primordial fluctuations. In the non-linear regime, the evolved $P(\delta)$ depends on the initial power spectrum and can serve as a probe of initial conditions (Bernardeau 1992; Juskiewicz, Bouchet & Colombi 1993). One can also extract moments of the full probability distribution. One interesting combination is the ratio of the third moment to the square of the second moment $S \equiv \langle \delta^3 \rangle / \langle \delta^2 \rangle^2$. For Gaussian initial conditions, second-order perturbation theory predicts that this ratio should remain constant. For a model with skewed initial probability distribution, $S$ will evolve and even change sign. The PDF is straightforward to measure from a galaxy survey. The principal complications are systematic distortions introduced by shot noise and peculiar velocities.

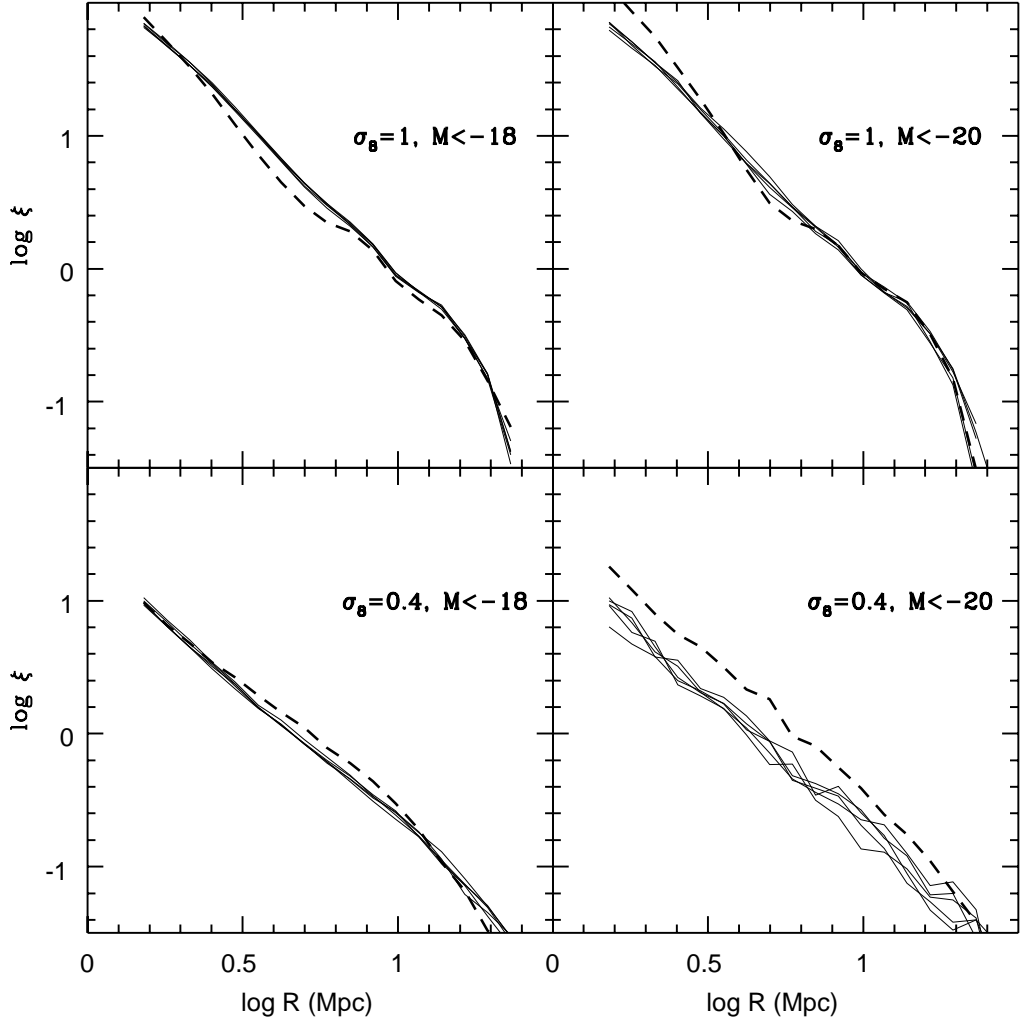

Figure 11: The two-point correlation function of galaxies is compared to that of dark matter. The dashed line in each panel is the galaxy correlation function. The thin solid lines are the correlation functions of random dark matter catalogues sampled at the same mean density as each galaxy catalogue.

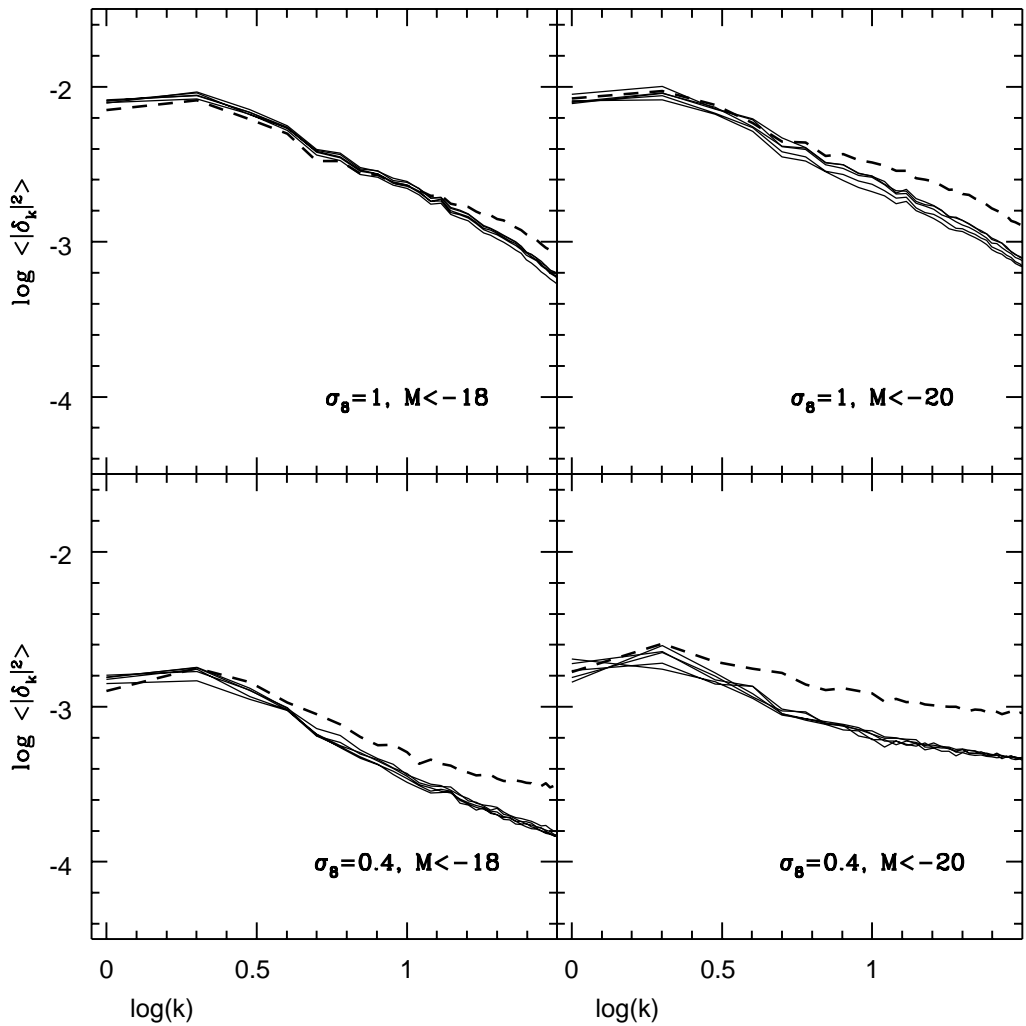

Figure 12: The power spectrum of galaxies is compared to that of dark matter. The explanation of the lines is as given in figure 11.

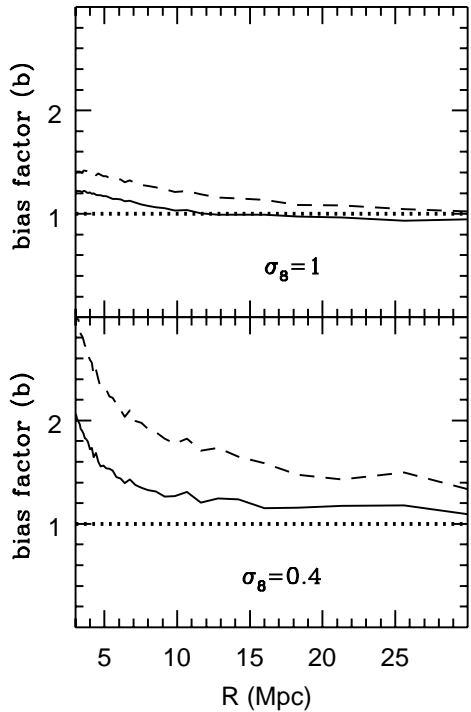

Figure 13: The linear bias factor as a function of radius evaluated from the power spectra shown in figure 12. The solid line is for galaxies with $M(B) \leq -18$. The dashed line is for galaxies with $M(B) \leq -20$.

The one-point probability distribution functions for our galaxy and dark matter samples are shown in figure 15. Both the galaxies and the dark matter have been smoothed with a Gaussian window function of radius 500 km s$^{-1}$. We have chosen to rescale the density field of the galaxies and the dark matter to the same fluctuation "level" by plotting $P(\delta/\sigma)$ against $\delta/\sigma$, where $\sigma$ is the rms fluctutation of the galaxies or the dark matter at this smoothing scale. The agreement between $P(\delta/\sigma)$ for galaxies and dark matter in both models is remarkable. There is agreement for both bright galaxies and faint galaxies. This is to be expected, as we have shown that a *linear* bias relation is a good approximation in our models and $P(\delta/\sigma)$ is not sensitive to linear bias. The PDF and its moments thus emerge as robust statistics for probing the properties of the initial density field.

# 4  Summary and Discussion

We have introduced a new method of calculating the bias in the distribution of galaxies relative to that of dark matter on large scales. Dark matter halos are identified as the sites where gas will be able to cool and then turn into stars to make a galaxy. The bias of halos relative to dark matter can be specified analytically, using an extension of the Press-Schechter theory, or by using N-body simulations of the evolution of the dissipationless dark matter component. The luminosities and morphologies of the galaxies contained within a dark matter halo of given mass are then specified using the semi-analytic models of Kauffmann, White & Guiderdoni (1993).

We demonstrate that the main factor that determines galaxy bias is the assumed normalization of the initial density fluctuation field, as this sets the value of $M_*$ at z=0. For models with $\sigma_8 \sim 1$, $M_* \sim 10^{14} M_\odot$, the mass of a typical cluster, and galaxies are on the whole unbiased with respect to the dark matter. For models with $\sigma_8 = 0.4$, $M_*$ is $\sim 10^{13} M_\odot$, the mass of a typical galaxy halo. Galaxies with luminosities less than $L_*$ are unbiased tracers of the mass, but more luminous galaxies are positively biased. Galaxies are thus predicted to be segregated by luminosity. More bias may be introduced by altering the shape of the CDM spectrum, so that it has more power on supercluster scales and less power on galaxy scales. This type of spectrum required in any case to fit the measured shape of the galaxy correlation function (Efstathiou, Bond & White 1992).

The galaxy bias relation will also differ according to the morphological type, colour or star formation rates of the galaxies selected to appear in the sample. Clusters are dominated by red elliptical galaxies that have typically not undergone recent star formation. Low mass halos contain blue disk systems with star formation rates of a few solar masses per year. Thus by selecting by type or by colour, one introduces additional biases towards either low or high density regions of the universe.

For all cosmological models and all luminosities and types of galaxies, the bias relation on large scales is to a good approximation *linear*. We have demonstrated that this a robust result and that it does not depend on the parameters of our galaxy formation model, such as star formation efficiency, feedback efficiency or merging rates.

Our analytic results are confirmed when we study the difference between the clustering properties of galaxies and dark matter directly, using galaxy catalogues constructed from N-body simulations. We also apply a number of different clustering statistics, including the correlation function, the power spectrum, the void probability function (VPF) and the one-point probability distribution function (PDF), to both the galaxies and the dark matter and see whether there are notable differences in the results. The PDF turns out to be a particularly robust statistic as no difference at all is found for galaxies and dark matter. The VPF, on the other hand, can give very different results for galaxies and dark matter, depending on the properties of the galaxy sample, and we conclude that its use is limited.

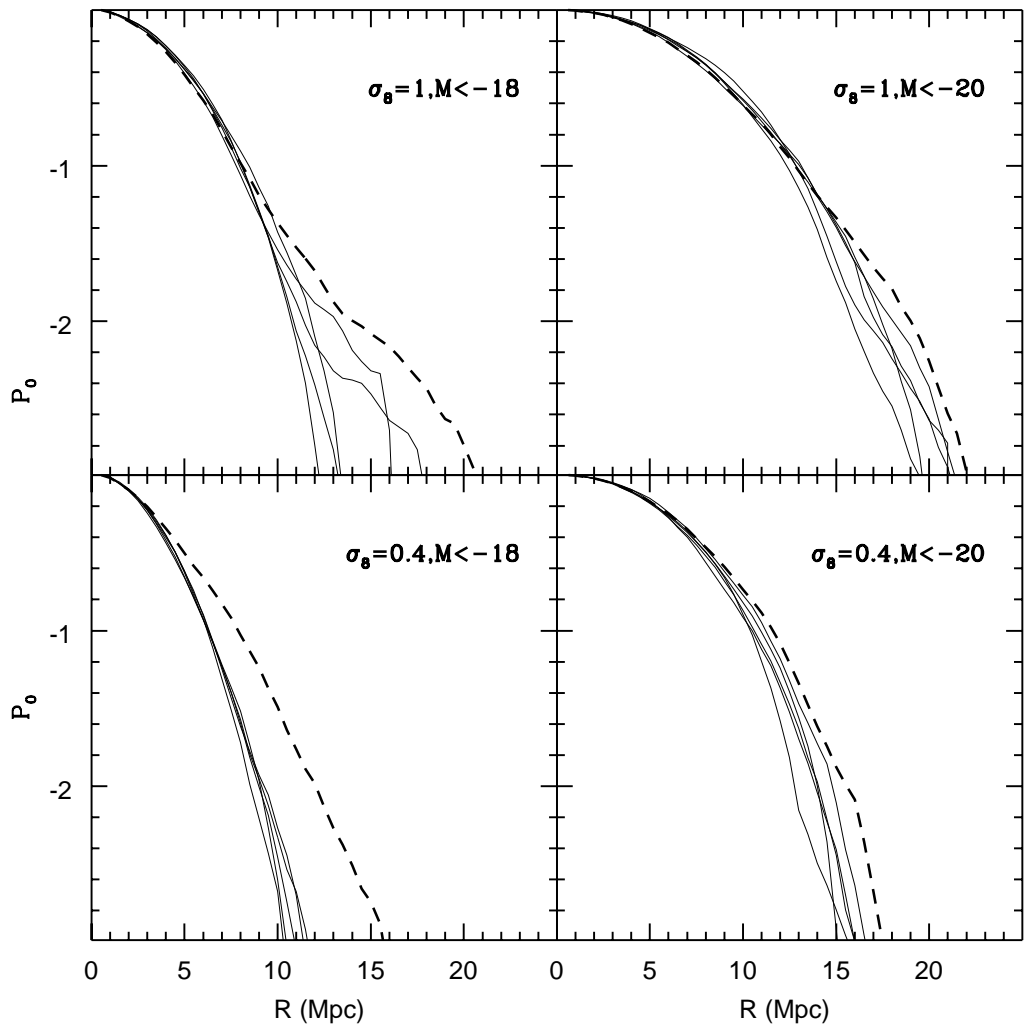

Figure 14: The void probability function of galaxies is compared to that of dark matter.

The weak link in our analysis is without doubt the halo-galaxy connection. In the future, it will be possible to use observational data to specify this connection, thereby bypassing the uncertainties associated with any theoretical model. Ramella, Geller & Huchra (1992) and Moore, Frenk & White (1993) have discussed methods for extracting catalogues of galaxy groups from redshift surveys. They also make dynamical estimates of the masses of their groups and use these to estimate $M_*$. The upcoming era of million-galaxy redshift surveys will permit such analyses to be carried out much more accurately, and for a much bigger sample of groups and clusters. It will also be possible to determine luminosity functions for groups and clusters down to fainter limiting magnitudes. More accurate cluster mass estimates will be obtained using techniques based on gravitational lensing (Kaiser & Squires 1993; Seitz & Schneider 1995). Future redshift surveys will thus present us with the possibility of *reconstructing* the underlying matter density distribution directly from the distribution of galaxies. We intend to explore this in more detail in a future paper.

## Acknowledgments

We acknowledge the hospitality of the Institute for Theoretical Physics, Santa Barbara where much of the work on this paper was carried out. AN acknowledges the support of a PPARC postdoctoral fellowship.

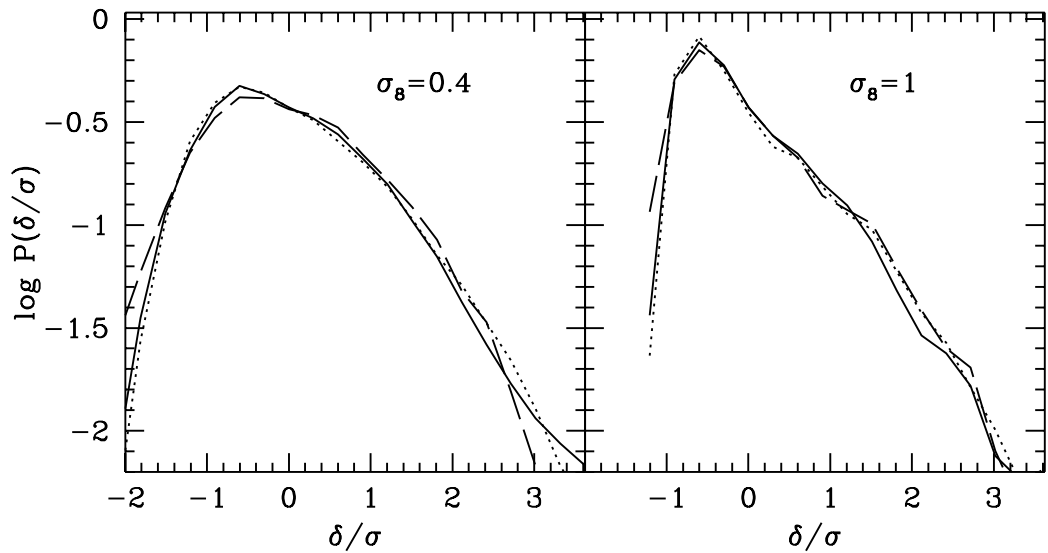

Figure 15: The one-point probability density function of galaxies is compared to that of dark matter. The solid line in each plot is $P(\delta/\sigma)$ for the dark matter, where $\sigma$ is the rms overdensity at a scale 500 km s$^{-1}$. The dotted line is for galaxies with $M(B) \leq -18$. The dashed line is for galaxies with $M(B) \leq -20$.